\documentclass{IEEETran}
 \usepackage{amsmath,amsthm,amssymb,amsfonts}
 \usepackage{hyperref}
 \usepackage{amssymb}
 \usepackage{array,multirow,graphicx}
 \usepackage{float}
 \usepackage{balance}
 
 \title{Thank you for Attention: A survey on Attention-based Artificial Neural Networks for Automatic Speech Recognition}
 
 \author{\IEEEauthorblockN{Priyabrata Karmakar, Shyh Wei Teng, Guojun Lu} 
 
\thanks{Priyabrata Karmakar, Shyh Wei Teng and Guojun Lu are with the School of Engineering, IT and Physical Sciences, Federation University Australia. e-mail:\{p.karmakar, shyh.wei.teng, guojun.lu\}@federation.edu.au}
}

\begin{document}
	
\maketitle	

\begin{abstract}
Attention is a very popular and effective mechanism in artificial neural network-based sequence-to-sequence models. In this survey paper, a comprehensive review of the different attention models used in developing automatic speech recognition systems is provided. The paper focuses on the development and evolution of attention models for offline and streaming speech recognition within recurrent neural network- and Transformer- based architectures.
\end{abstract}

	
\section{Introduction}

Automatic speech recognition (ASR) is a type of sequence-to-sequence (seq2seq) task. The input speech sequence is transcribed into a sequence of symbols.  The majority of the existing state-of-the art ASR systems consisted of three modules: acoustic, pronunciation and language \cite{jelinek1976continuous}. These three modules are separately trained. The acoustic module predicts phonemes based on the input speech feature like Mel Frequency Cepstral Coefficient (MFCC) \cite{muda2010voice}. The pronunciation module is a hidden Markov model \cite{gales2008application} which  maps the phonemes predicted at the earlier module to word sequences. Finally, the language module which is pre-trained on a large corpus, scores the word sequences. In other words, language model estimates the probabilities of next word based on previously predicted words to establish a meaningful sentence. This traditional approach has some limitations. First, the modules are trained separately for different objective functions. Therefore, it may result incompatibility between modules. Also separate training is time expensive. Second, the pronunciation model requires a dictionary for mapping between phonemes and word sequences. The pronunciation dictionary is developed by linguistic experts and is prone to human errors \cite{kudo2004applying, bird2006nltk}.

From the last decade, deep learning has been applied significantly in various domains, such as image and video processing, machine translation and text processing. Speech recognition is not an exception as well. Early deep learning-based ASR systems mostly consider a hybrid approach where the acoustic model is replaced by a deep neural network and the rest of modules use the traditional approach \cite{mohamed2009deep, hinton2012deep, graves2013hybrid}.

The recent trend of building ASR systems is to develop an end-to-end deep neural network. The network can therefore map the input speech sequence to a sequence of either graphemes, characters or words. In end-to-end ASR systems, the acoustic, pronunciation and language modules are trained jointly to optimize a common objective function and the network overcomes the limitations of traditional ASR systems. In the literature, there are generally two major end-to-end ASR architectures can be found. They are (a) Connectionist temporal classification (CTC)-based, and (b) Attention-based. CTC uses Markov assumptions to solve sequence-to-sequence problem with a forward-backward algorithm \cite{ctc}. Attention mechanism aligns the relevant speech frames for predicting symbols at each output time step \cite{chorowski2014end,chorowski2015attention}.

The end-to-end ASR models are mainly based on an encoder-decoder architecture. The encoder part converts the speech frames and their temporal dependencies into a high level representation which will be used by the decoder for output predictions. The initial versions of the encoder-decoder architecture for ASR modelled with recurrent neural network (RNN) as the main component for sequence processing \cite{graves2012sequence, graves2013speech}. RNN is a type of artificial neural network which is typically used for modelling sequential data. Apart from the vanilla RNN, some other variations like long short-term memory (LSTM) \cite{hochreiter1997long}, gated recurrent unit (GRU) \cite{cho2014learning} are also popular in modelling sequential data. RNNs can be used in unidirectional as well as bi-directional fashion \cite{schuster1997bidirectional, graves2005bidirectional}. Convolutional neural networks (CNN) coupled with RNNs \cite{7953077} or stand-alone \cite{zhang2016towards} have also been used to make effective ASR models. Processing data sequentially is an inefficient process and may not capture temporal dependencies effectively. To address the limitations of RNN, Transformer network \cite{vaswani2017attention} has been recently proposed for sequence-to-sequence transduction. Transformer is a recurrence-free encoder-decoder architecture where sequence tokens are processed parallelly using self-attention mechanism.

Automatic speech recognition operates in two different modes: offline (when recorded speech is available before transcription starts), and online or streaming (when transcription starts simultaneously as the speaker(s) starts speaking). In this paper, we have reviewed attention-based ASR literature for both offline and streaming speech recognition. While reviewing, we have only considered the models built with either recurrent neural network (RNN) or Transformer. Nowadays, ASR models are widely embedded in systems like smart devices and chatbots. In addition, application of attention mechanism is showing great potential in achieving higher effectiveness and efficiency for ASR. From the middle of last decade, a lot of progress has been made on attention-based models. Recently, some survey papers \cite{chaudhari2019attentive, galassi2020attention} have presented the development of attention-based models on natural language processing (NLP). These survey papers have documented the advancement of a wide range of NLP applications like machine translation, text and document classification, text summarisation, question answering, sentiment analysis, and speech processing.  However, the existing literature still lacks a survey specifically targeted on the evolution of attention-based models for ASR. Therefore, we have been motivated to write this paper.

The rest of paper is organised as follows. Section \ref{sec:attention} provides a simple explanation of Attention mechanism.  A brief introduction to attention-based encoder-decoder architecture is discussed in Section \ref{sec:attention-aed}. Section \ref{sec:offline} discusses the evolution of offline speech recognition followed by the evolution of streaming speech recognition in Section \ref{sec:streaming}. Finally Section \ref{sec:conclusion} concludes the paper.

	\begin{table}
		\centering
		\caption{Different types of attention mechanism for ASR }
		\label{attention_types}
		\begin{tabular}{|>{\centering}m{1.2in}|  >{\arraybackslash}m{1.7in} |}
			\hline
		Name &   \; \; \; \; \; \; \; Short description    \\ \hline
		Global/Soft \cite{chorowski2014end}& At each decoder time step, all encoder hidden states are attended.\\ \hline
		Local/Hard \cite{bahdanau2016end} &  At each decoder time step, a set of encoder hidden states (within a window) are attended.\\ \hline
		Content-based \cite{chan2016listen} & Attention calculated only using the content information of the encoder hidden states.\\ \hline
		Location-based \cite{meng2019character} & Attention calculation depends only on the decoder states and not on the encoder hidden states. \\ \hline
		Hybrid \cite{chorowski2015attention} & Attention calculated using both content and location information. \\ \hline
		Self \cite{vaswani2017attention}& Attention calculated over different positions(or tokens) of a sequence itself. \\ \hline
		2D \cite{dong2018speech} & Attention calculated over both time- and frequency-domains. \\ \hline
		Hard monotonic  \cite{raffel2017online} & At each decoder time step, only one encoder hidden state is attended.\\ \hline
		Monotonic chunkwise \cite{chiu2018monotonic} & At each decoder time step, a chunk of encoder states (prior to and including the hidden state identified by the hard monotonic attention) are attended. \\ \hline
		Adaptive monotonic chunkwise  \cite{fan2019online} & At each decoder time step, the chunk of encoder hidden states to be attended is computed adaptively.\\ \hline

			\hline
		\end{tabular}
	\end{table}

	\section{Attention}
	
	\label{sec:attention}
	
	Attention mechanism can be defined as the method for aligning relevant frames of input sequence for predicting the output at a particular time step. In other words, attention mechanism helps deciding which input frame(s) to be focused at and how much for the output prediction at the corresponding time step. With the help of a toy example, the attention mechanism for sequence-to-sequence model is explained in this section. Consider the input source sequence is $ X $ and the output target sequence is $ Y $. For simplicity, we have considered the number of frames (or tokens) in both input and output sequence is same.
	
	 \begin{equation*}
	 X = [x_1, x_2,\cdots ,x_n]  ; 	 Y = [y_1, y_2,\cdots ,y_n]. 
	 \end{equation*}
	
	An encoder processes X to a high level representation (hidden states) and passes it to the decoder where prediction of Y happens. In most cases, the information required to predict a particular frame $ y_t $ is confined within a small number of input  frames. Therefore, for decoding $ y_t $, it is not required to look at each input frames. The Attention model aligns the input frames with $ y_t $ by assigning match scores to each pair of input frame and $ y_t $. The match scores convey how much a particular input frame is relevant to $ y_t $ and accordingly, the decoder decides the degree of focus on each input frame for predicting $ y_t $.
	
	Depending on how the alignments between output and input frames are designed, different types of attention mechanism are presented in the literature. A list of existing attention models along with short descriptions is provided in Table \ref{attention_types}. The detailed explanation of different attention models is discussed throughout the paper. In this survey, we have considered the models which are built within RNN or Transformer architecture. Table \ref{literature_types} provides the list of literature which we have reviewed in the later sections of this paper.

		\begin{table}
		\centering
		\caption{List of literature }
		\label{literature_types}
		\begin{tabular}{|>{\centering}m{0.8in}|>{\centering}m{0.8in}| >{\centering\arraybackslash}m{0.8in} |}
			\hline
			Attention & Offline ASR & Streaming ASR  \\ \hline
			RNN-based & \cite{chorowski2014end,chorowski2015attention, chan2016listen, zeyer2018improved, meng2019character, bahdanau2016end, chan2016online, tjandra2017local, kim2017joint, watanabe2017hybrid, hori2017advances, watanabe2017language, moritz2019triggered} & \cite{hou2017gaussian, raffel2017online, liu2020multi, chiu2018monotonic, kim2019attention, fan2019online, jaitly2016online, sainath2018improving, wang2020attention, inaguma2020minimum, hsiao2020online}\\ \hline
			Transformer-based & \cite{dong2018speech, zhou2018syllable, zhou2018comparison, li2019speechtransformer, pham2019very, irie2019language, irie2020much, wang2020transformer, sperber2018self, sperber2018self, katharopoulos2020transformers, dai2019transformer, rae2019compressive, zhang2020transformer, li2019improving, shi2020weak} & \cite{dong2019self, wu2020streaming, yeh2019transformer, mohamed2019transformers, zhang2020transformer, tian2019self, wang2020transformer, povey2018time, moritz2020streaming, tian2020synchronous, miao2020transformer, miao2019online, inaguma2020enhancing}  \\

				\hline
\end{tabular}
\end{table}

	\section{Attention-based Encoder-Decoder}
	
	\label{sec:attention-aed}
	
	 For ASR, attention-based encoder-decoder architecture is broadly classified into two categories: (a) RNN-based, and (b) Transformer-based. In this section, we have provided an overview of both categories. In the following sections, 
	 a detailed survey has been provided.
	
	\subsection{RNN-based encoder-decoder architecture}
	Sequence-to-sequence RNN-based ASR models are based on an encoder-decoder architecture. The encoder is an RNN which takes input sequence and converts it into hidden states. The decoder is also an RNN which takes the last encoder hidden state as input and process it to decoder hidden states which in turn used for output predictions. This traditional encoder-decoder structure has some limitations:
	
	\begin{itemize}
		\item The encoder hidden state, $ h_T $ (last one) which is fed to the decoder has the entire input sequence information compressed into it. For longer input sequences, it may cause information loss as $ h_T $ may not capture long-range dependencies effectively.
		
		\item There is no alignment between the input sequence frames and the output. For predicting each output symbol, instead of focusing on the relevant ones, the decoder considers all input frames with same importance.
	\end{itemize}
	
The above issues can be overcome by letting the decoder to access all the encoder hidden states (instead of the last one) and at each decoder time step, relevant input frames are given higher priorities than others. It is achieved by incorporating attention mechanism to the encoder-decoder model. As a part of sequence-to-sequence modelling, attention mechanism was introduced in \cite{bahdanau2015neural} for machine translation. Inspired by the effectiveness in \cite{bahdanau2015neural}, the attention mechanism was introduced to ASR in \cite{chorowski2015attention}. An earlier version of this work has been presented in \cite{chorowski2014end}.

\label{sec:global}

\begin{figure}
	\centering
	\includegraphics[width=0.8\linewidth]{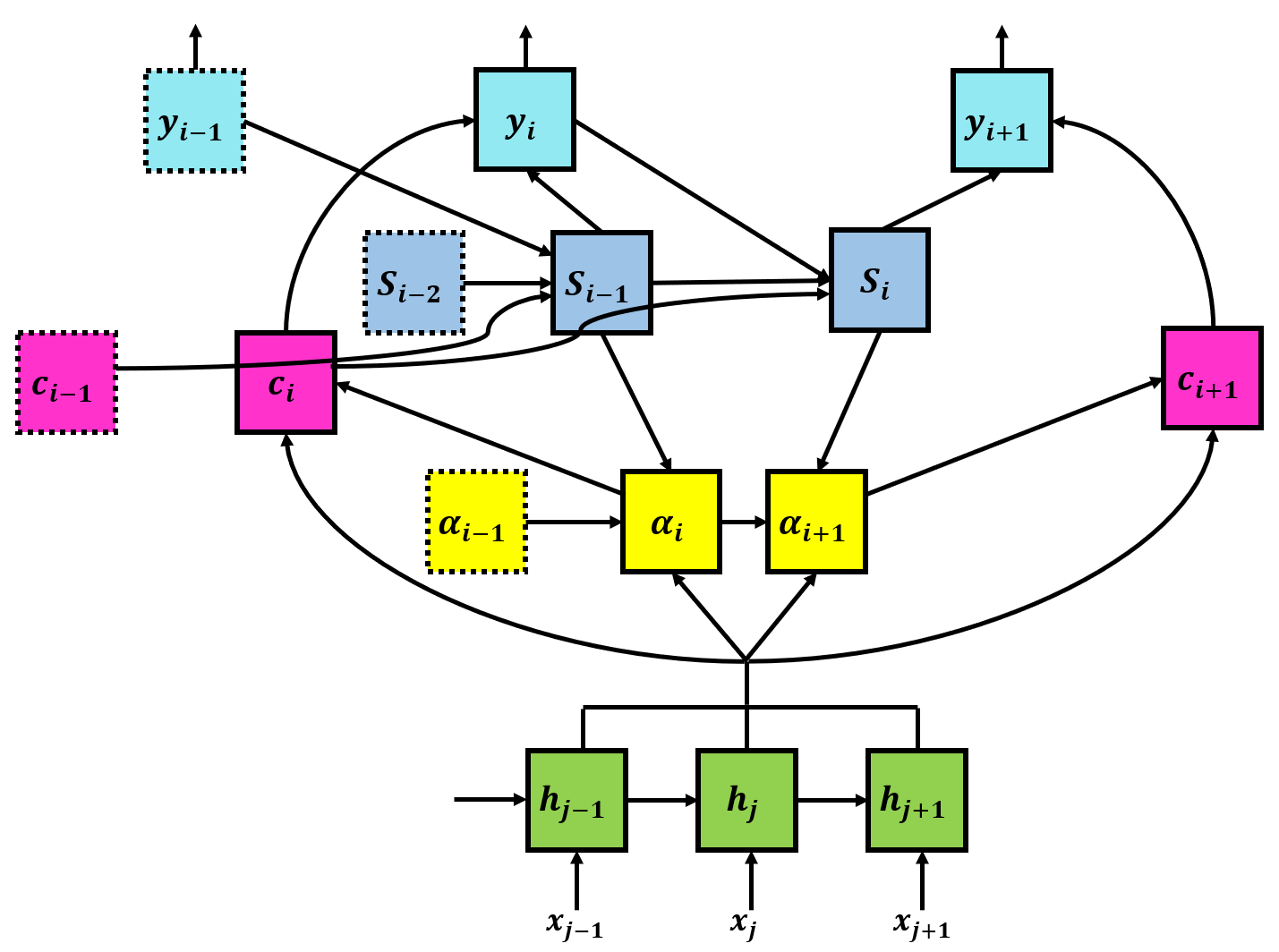}
	\caption{RNN-based encoder-decoder architecture with attention}
	\label{fig:rnn-attention}
\end{figure}

The model in \cite{chorowski2015attention} is named as attention-based recurrent sequence generator (ASRG). The graphical representation of this model is shown in Figure \ref{fig:rnn-attention}. The encoder of ASRG processes the input audio frames to encoder hidden states which are then used to predict output phonemes. By focusing on the relevant encoder hidden states, at $ i^{th} $ decoder time step, prediction of phoneme $ y_i $ is given by \eqref{eq:spell}

\begin{equation}
y_i = Spell(s_{i-1}, c_i)
\label{eq:spell},
\end{equation}
where $ c_i $ is the context given by \eqref{eq:context} generated by attention mechanism at the $ i^{th} $ decoder time step.  $ s_i $ given by \eqref{eq:decoder-state} is the decoder hidden state at $ i^{th} $ time step. It is the output of a recurrent function like LSTM or GRU. $ Spell(.,.) $ is a feed-forward neural network with softmax output activation.

\begin{equation}
c_i = \sum_{j=1}^L\alpha_{i,j}h_j 
\label{eq:context},
\end{equation} where $ h_j $ is the encoder hidden state at the $ j^{th} $ encoder time step. $ \alpha_{i,j} $ given by \eqref{eq:alpha} is the attention probability belonging to the $ j^{th} $ encoder hidden state for the output prediction at $ i^{th} $ decoder time step. In other words, $ \alpha_{i,j} $ captures the importance of the $ j^{th} $ input speech frame (or encoder hidden state) for decoding the $ i^{th} $ output word (or phoneme or character). $ \alpha_i $ values are also considered as the alignment of encoder hidden states ($ h_{j\in [1,\dotsm, L]} $) to predict an output at $ i^{th} $ decoder time step. Therefore, $ c_i $ is the sum of the products (SOP) of attention probabilities and the hidden states belonging to all encoder time steps at the $ i^{th} $ decoder time step and it provides a context to the decoder to decode (or predict) the corresponding output.

\begin{equation}
s_i = \textit{Recurrent}(s_{i-1}, c_i, y_{i-1}).
\label{eq:decoder-state}
\end{equation}


\begin{equation}
\alpha_{i,j}=\dfrac{exp(e_{i,j})}{\sum_{j=1}^L exp(e_{i,j}) },
\label{eq:alpha}
\end{equation} where $ e_{i,j} $ is the matching score between the $ i^{th} $ decoder hidden state and the $ j^{th} $ encoder hidden state. It is computed using a hybrid attention mechanism given by \eqref{eq:matchg} in a general form and by \eqref{eq:matchp} in a parametric form.

\begin{equation}
e_{ij} = Attend(s_{i-1}, \alpha_{i-1}, h_j).
\label{eq:matchg}
\end{equation}

\begin{equation}
e_{i,j} = w^Ttanh(Ws_{i-1}+Vh_j+Uf_{i,j}+b),
\label{eq:matchp}
\end{equation}
where $ w $ and $ b $ are vectors and $ W $, $ V $ and $ U $ are matrices. These are all trainable parameters. $ f_i = F \ast \alpha_{i-1} $ is a set of vectors which are extracted for every encoder state $ h_j $ of the previous alignment $ \alpha_{i-1} $ which is convolved with a trainable matrix $ F $. The $ tanh $ function produces a vector. However, $ e_{i,j} $ is a single score. Therefore, a dot product of $ tanh $ outcome and $ w $ is performed.
The mechanism in \eqref{eq:matchg} is referred to as hybrid attention as it considers both location ($ \alpha $) and content ($ h $) information. By dropping either $ \alpha_{i-1} $ or $ h_j $, the $ Attend $ mechanism is called content-based or location-based attention.


\subsection{Transformer-based encoder-decoder architeture}
RNN-based encoder-decoder architecture is sequential in nature. To capture the dependencies, hidden states are generated sequentially and at each time step, the generated hidden state is the output of a function of previous hidden state. This sequential process is time consuming. Also, during the training, error back propagates through time and this process is again time consuming.

To overcome the limitations of RNN, Transformer network is proposed completely based on attention mechanism. In Transformer network, no recurrent connection is used. Instead, the input farmes are processed parallelly at the same time, and during training, no back propagation through time is applicable.

\begin{figure}
	\centering
	\includegraphics[width=1\linewidth]{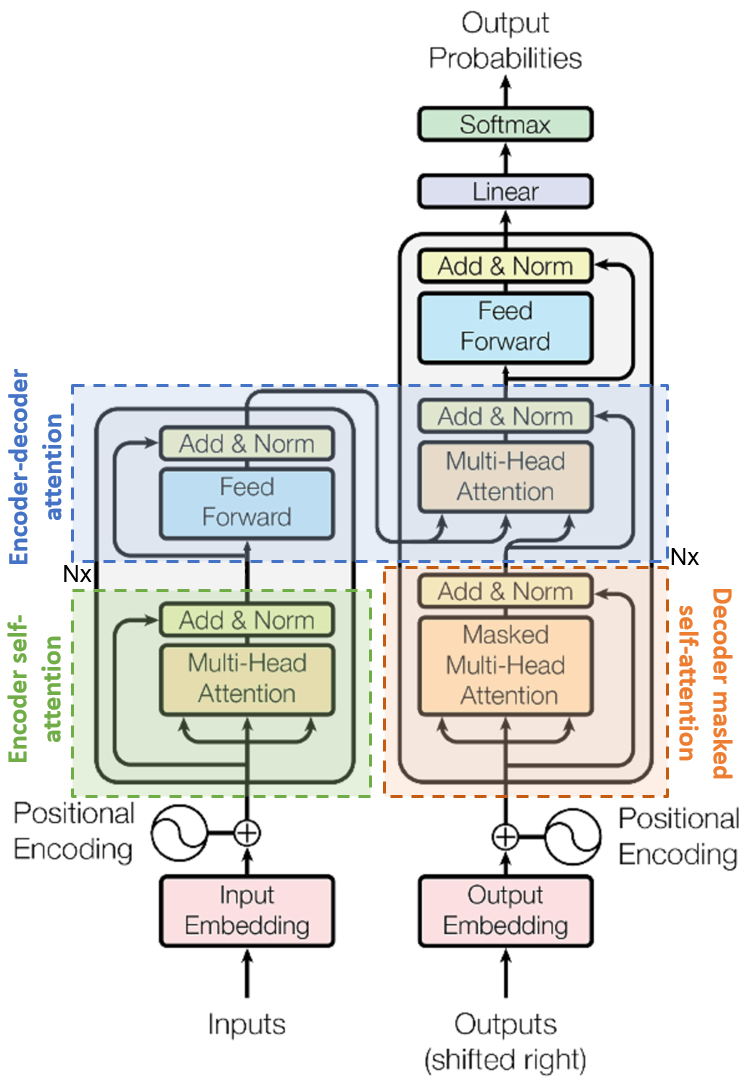}
	\caption{Transformer-based encoder-decoder architecture \cite{vaswani2017attention}}
	\label{fig:transformer}
\end{figure}

Transormer network was introduced in \cite{vaswani2017attention} for machine translation and later it is successfully applied to ASR tasks. In this section, the idea of Transformer is given as described in \cite{vaswani2017attention}. The graphical representation of Transformer is shown in Figure \ref{fig:transformer}.

The Transformer network is composed of an encoder- decoder architecture but there is no recurrent or convolutional neural network involved here. Instead, the authors have used self-attention to incorporate the dependencies in the seq2seq framework. The encoder is composed of six identical layers where each layer is divided into two sub-layers. The first sub-layer is a multi-head self-attention module and the second one is a position-wise feed-forward neural network. The decoder is also composed of six identical layers but has an additional sub-layer to perform multi-head self-attention over the encoder output. Around each sub-layer, a residual connection \cite{he2016deep} is employed followed by a layer-normalisation \cite{ba2016layer}. In the decoder section, out of two  multi-head attention blocks, the first one is masked to prevent positions from attending subsequent positions. 

The attention function is considered here as to obtain an output which is the weighted sum of values based on matching a query with keys from the corresponding key-value pairs using scaled dot-product. The dimensionalities of query, key and value vectors are $ d_k $, $ d_k $ and $ d_v $, respectively. In practice, attention is computed on a set of query, key and value together by stacking these vectors in a matrix form. Mathematically, it is given by \eqref{eq:self-attention}.

\begin{equation}
Attention(Q,K,V) = Softmax(\frac{QK^T}{\sqrt(d_k)})V
\label{eq:self-attention}, 
\end{equation} where $ Q $, $ K $, $ V $ are matrices which represent Query, Key and Value, respectively.

Positional information is added to the input sequence to generate the input embedding upon which the attention will be performed. Instead of directly applying attention on input embeddings, they are linearly projected to $ d_k $ and $ d_v $ dimensional vectors using learned projections given by \eqref{eq:attention-projection}

\begin{equation}
\begin{split}
q = X W_q, \\
k = X W_k, \\
v = X W_v,
\end{split}
\label{eq:attention-projection}
\end{equation} where $W_q \in \mathcal{R}^{d_{model}\times d_k}$, $W_k \in \mathcal{R}^{d_{model}\times d_k}$ and $W_v \in \mathcal{R}^{d_{model}\times d_v}$ are trainable parameters. $ d_{model} $ is the dimension of input embeddings. $ X $ is the input embedding for the encoder section and the output embedding for the masked multi-head block for the decoder section. For the second multi-head block of the decoder section, $ X $ is the encoder output for $ k $ and $ v $ projection. However, for $ q $ projection, $ X $ is the output from the masked multi-head section.

In Transformer network \cite{vaswani2017attention}, the attention mechanism have been used in three different ways. They are as follows.

\begin{enumerate}
	\item Encoder self-attention: In the encoder section, attention mechanism is applied over the input sequences to find the similarity of each token of a sequence with rest of the tokens.
	\item Decoder masked self-attention: Similar to the encoder self-attention, output (target) sequence tokens attend each other in this stage. However, instead of accessing the entire output sequence at a time, the decoder can only access the tokens preceding the token which decoder attempts to predict. This is done by masking current and all the future tokens of a particular decoder time step. This approach prevents the training phase to be biased.
	
	\item Encoder-decoder attention: This occurs at the decoder section after decoder masked self-attention stage. With reference to \eqref{eq:self-attention}, at this stage, Q is the linear projection of the vector coming from decoder's masked self-attention block. Whereas, K and V are obtained by linearly projecting the vector resulting from encoder self-attention block. This is the stage where the mapping between input and output (target) sequences happens. The output of this block is the attention vectors containing the relationship between tokens of input and output sequences.
\end{enumerate}

At each sub-layer, the attention is performed $ h $-times in parallel. Hence, the name ``multi-head attention" is given. In \cite{vaswani2017attention}, the value of $ h $ is 8. According to the authors, multi-head attention allows the model to jointly attend to information from different representation
subspaces at different positions. The outputs from each attention head are then concatenated and projected using \eqref{eq:multihead} to obtain the final output of the corresponding sub-layer.  

\begin{equation}
MultiHead(Q,K,V) = Concat(head_i, \dotsm, head_h)W_o,
\label{eq:multihead}
\end{equation}

where $ head_{i \in [1,h]} $ is computed using \eqref{eq:attention-projection} and $W_o \in \mathcal{R}^{hd_v \times d_{model}}$ is a trainable parameter.

\section{Offline Speech Recognition}

\label{sec:offline}

In this section, the evolution of attention-based models will be discussed for offline speech recognition. This section is divided into four sub-sections to explore global and local attention with RNN-based models, joint attention-CTC with RNN-based models and RNN-free Transformer-based models.

\subsection{Global Attention with RNN}

Global attention is computed over the entire encoder hidden states at every decoder time step. The mechanism illustrated in Section \ref{sec:global} as per \cite{chorowski2015attention} is an example of global attention. Since \cite{chorowski2015attention}, a lot of progress has been made by many researchers.

 The authors of \cite{chan2016listen} presented a global attention mechanism in their   \textit{Listen, Attend and Spell }(LAS)  model. Here, $ Spell $ function takes inputs as current decoder state $ s_i $ and the context $ c_i $. $ y_i = Spell(s_i, c_i) $.  $ s_i $ is computed using a recurrent function which takes inputs as previous decoder state ($ s_{i-1} $), previous output prediction ($ y_{i-1} $) and previous context ($ c_{i-1} $). $s_i = Recurrent(s_{i-1}, y_{i-1}, c_{i-1})$. The authors have used the content information only to calculate the matching scores given by \eqref{eq:attentionc}. Attention probabilities are then calculated by \eqref{eq:alpha} using the matching scores.
 
\begin{equation}
e_{i,j} = w^Ttanh(Ws_{i-1}+Vh_j+b).
\label{eq:attentionc}
\end{equation} A similar content-based global attention have been proposed in \cite{zeyer2018improved} where a feedback factor is incorporated in addition to the content information in calculating the matching scores for better numerical stability. In generalised form, it is given by \eqref{eq:attentionf}

\begin{equation}
e_{i,j} = w^T tanh (W[s_i, h_j, \beta_{i,j}])
\label{eq:attentionf},
\end{equation} where $ \beta_{i,j} $ is the attention weight feedback computed using the previously aligned attention vectors and it is given by \eqref{eq:weightf}.

\begin{equation}
\beta_{i,j} = \sigma(w_b^Th_j)\cdot\sum_{k=1}^{i-1}\alpha_{k,j}
\label{eq:weightf},
\end{equation} where $ w_b $ is a trainable weight vector. Here, $ Spell $ function is computed over $ s_i $, $ y_{i-1} $ and $ c_i $, i.e.  $ y_i = Spell(s_i, y_{i-1},c_i) $

	 A character-aware (CA) attention is proposed in \cite{meng2019character} to incorporate morphological relations for predicting words and sub-word units (WSU). A separate RNN (named as CA-RNN by the author) which dynamically generates WSU representations connected to the decoder in parallel with the encoder network. The decoder hidden state $ s_{t-1} $ is required to obtain the attention weights at $ t $ time step. $ s_t $ is computed using the recurrent function over  $ s_{t-1} $, $ w_{t-1} $(WSU represenation) and $ c_{t-1} $. The matching scores required to compute attention vectors at decoder $ t $ time step is calculated using \eqref{eq:matchp}. In contrast to \cite{chorowski2015attention}, the authors have used $ RELU $ instead of $ tanh $ function and claimed it provides better ASR performance.

	\subsection{Local attention with RNN}

	In global attention model, each encoder hidden states are attended at each decoder time step. This results in a quadratic computation complexity. In addition, the prediction of a particular decoder output mostly depends on a small number of encoder hidden states. Therefore, it is not necessary to attend the entire set of encoder hidden states at each decoder time step. The application of local attention fulfils the requirement of reducing the computation complexity by focusing on relevant encoder hidden states. Local attention mechanism is mostly popular in streaming speech recognition but, it has been applied to offline speech recognition as well. The core idea of local attention is to attend a set of encoder hidden states within a window or range at each decoder time step instead of attending the entire set of encoder hidden states. Local attention was introduced in \cite{effective} for machine translation and thereafter, it has been applied to ASR as well.

	In \cite{bahdanau2016end}, the window upon which the attention probabilities are computed is considered as $ [m_{t-1}-w_l, m_{t-1}+w_r] $, where $ m_{t-1} $ is the median of previous alignment $\alpha_{t-1}$ (i.e. the attention probabilities computed at the last decoder time step). $ w_l $ and $ w_r $ are the user-defined fixed parameters which determine the span of the window in left and right directions, respectively. A similar local attention was proposed in \cite{chan2016online}.
	
To obtain the attention window, position difference $ \bigtriangleup p_t $ is calculated for the prediction at the $ t $ decoder time step in \cite{tjandra2017local}. $ \bigtriangleup p_t $ is the position difference between the centre of attention windows of previous and current decoder time steps. Therefore, given  $ p_{t-1} $ (the centre of previous attention window) and $ \bigtriangleup p_t $, the centre of current attention window can be calculated. After that, the attention window at the $ t^{th} $ decoder time step is set as $ [p_t -  \bigtriangleup p_t, p_t +  \bigtriangleup p_t ] $. Two methods were proposed to estimate $ \bigtriangleup p_t $ as given by \eqref{eq:constarined} and \eqref{eq:unconstarined}.
	
	\begin{equation}
	\bigtriangleup p_t = C_{max}\ast sigmoid(V_P^T tanh (W_ph_t^d)),
	\label{eq:constarined}
	\end{equation}
	where $ V_p $ and $ W_p $ are a trainable vector and matrix respectively. $ C_{max} $ is a hyper parameter to maintain the condition: $ 0<\bigtriangleup p_t<C_{max} $.
	
	\begin{equation}
	\bigtriangleup p_t =  exp(V_P^T tanh (W_ph_t^d)),
	\label{eq:unconstarined}
	\end{equation}
	
	Equations \eqref{eq:constarined} and \eqref{eq:unconstarined} are named as Constrained and Unconstrained position predictions respectively.

	\subsection{{Joint attention-CTC with RNN}}

	Two main approaches for end-to-end encoder-decoder ASR are attention-based and CTC \cite{hannun2014deep}-based. In attention-based approach, the decoder network finds an alignment of the encoder hidden states during the prediction of each element of output sequence. The task of speech recognition is mostly monotonic. Therefore, the possibility of right to left dependency is significantly lesser compared to left to right dependency in ASR tasks. However, due to the flexible nature of attention mechanism, non-sequential alignments are also considered. Therefore, noise and irrelevant frames (encoder hidden states) may result in misalignment. This issue becomes worse for longer sequences as the length of input and output sequences vary due to factors, e.g. the rate of speech, accent, and pronunciation. Therefore, the risk of misalignment in longer sequences is higher. In contrast, CTC allows strict monotonic alignment of speech frames using forward-backward algorithm \cite{ctc,graves2014towards} but assumes targets are conditionally independent on each other. Therefore, temporal dependencies are not properly utilised in CTC, unlike in attention mechanism. For effective ASR performance, many researchers have combined the advantages of both attention and CTC in a single model and therefore, the CTC probabilities replaces the incorrect predictions by the attention mechanism.

	The discussion on CTC and its application on ASR is beyond the scope of this paper. However, in this section a brief introduction to CTC and how it is jointly used with attention is provided \cite{kim2017joint,watanabe2017hybrid}. CTC monotonically maps an input sequence to output sequence. Considering the model outputs $ L $- length letter sequence $Y \{y_l \in U|l = 1,\dotsm , L\}$ with a set of distinct characters $ U $, given the input sequence is $ X $. CTC introduces
	frame-wise letter sequence with an additional ``blank" symbol $ Z = \{z_t \in U \cup blank| t = 1, \dotsm, T    \} $. By using conditional independence
	assumptions, the posterior distribution $ p(Y|X) $ is factorized as follows:
	
	\begin{equation}
	 p(Y|X) \approx \underbrace {\sum_Z \prod_t p(z_t|z_{t-1},Y)p(z_t|X)p(Y)}_{ \triangleq p_{ctc}(Y|X)}.
	\end{equation}
	
	CTC has three distribution components by
	the Bayes theorem similar to the traditional or hybrid ASR. They are frame-wise posterior distribution $ p(z_t|X) $ - acoustic module, transition probability $ p(z_t|z_{t-1},C) $ - pronunciation module, and letter-based language module $ p(Y) $.

	Compared with CTC approaches, the attention-based approach does not make any conditional independence assumptions, and directly estimates the posterior $ p(Y|X) $ based on the chain rule:
	
	\begin{equation}
	p(Y|X) = \underbrace{\prod_l p(y_l|y_1, \dotsm, y_{l-1},X)}_{\triangleq p_{att}(Y|X)}.
	\end{equation}
	
	$ p_{ctc}(Y|X) $ and $ p_{att}(Y|X) $ are the CTC-based and attention-based objective functions, respectively. Finally, the logarithmic linear combination of CTC- and attention-based objective functions given by \eqref{eq:joint-objective} is maximised to leverage the CTC and attention mechanism together in a ASR model.
	
	
	\begin{equation}
	L = \lambda \: log \: p_{ctc}(Y|X) + (1 - \lambda)  \: log \: p_{att}(Y|X) 
	\label{eq:joint-objective},
	\end{equation} 	$ \lambda $ is a tunable parameter in the range $ [0,1] $.
	
	In \cite{kim2017joint, watanabe2017hybrid}, the CTC objective function was incorporated in the attention-based model during the training only. However, motivated by the effectiveness of this joint approach, in \cite{hori2017advances, watanabe2017language}, it is used for decoding or inferencing phase as well.

	A triggered attention mechanism is proposed in \cite{moritz2019triggered}. At each decoder time step, the encoder states which the attention model looks upon are controlled by a trigger model. The encoder states are shared with the trigger model which is a CTC-based network as well as with the attention model. The trigger sequence which is computed based on the CTC generated sequence provides alignment information that controls the attention mechanism. Finally, the objective functions of CTC and attention model are optimised jointly.

\subsection{RNN-free Transformer-based models}

Self-attention is a mechanism to capture the dependencies within a sequence. It allows to compute the similarity between different frames in the same sequence. In other words, self-attention finds to what extent different positions of a sequence relate to each other. Transformer network \cite{vaswani2017attention} is entirely built using self-attention for seq2seq processing and has been successfully used in ASR as well.

Transformer was introduced to ASR domain 
in \cite{dong2018speech} by proposing Speech-transformer. Instead of capturing only temporal dependencies, the authors of  \cite{dong2018speech} have also captured spectral dependencies by computing attention along time and frequency axis of input spectrogram features. Hence, this attention mechanism is named as ``2D attention". The set of $ (q,k,v) $ for time-domain attention is computed using \eqref{eq:attention-projection}. Here, the input embedding ($ X $) is the convolutional features of spectrogram. For frequency-domain attention, the set of $ (q,k,v) $ are the transpose of same parameters in the time-domain. At each block of multi-head attention, the time-domain and frequency-domain attentions are computed parallelly and after that they are concatenated using \eqref{eq:multihead}. In this case attention heads belong to both time and frequency domains. Speech transformer was built to output word predictions and later on it is explored for different modelling units like phonemes, syllables, characters in \cite{zhou2018syllable, zhou2018comparison} and for large-scale speech recognition in \cite{li2019speechtransformer}.

A very deep Transformer model for ASR is proposed in \cite{pham2019very}. The authors have claimed that depth is an important factor for obtaining effective ASR performance using Transformer network. Therefore, instead of using the original version of six stacked layers for both encoder and decoder, more layers (deep configuration) are used in the structure. Specifically, the authors have shown $ 36-12 $ layers for the encoder-decoder is the most effective configuration. To facilitate the training of this deep network, around each sub-layer, a stochastic residual connection is employed before the layer-normalisation. Another deep Transformer model is proposed in \cite{irie2019language} where it has been shown that the ASR performance is continually increased with the increase of layers up to 42 and the attention heads up to 16. The effect on performance beyond 42 layers and 16 attention-heads is not provided, probably due to the increased computation complexity. 
The authors have also experimentally shown that sinusoidal positional encoding \cite{vaswani2017attention} is not required for deep Transformer model. To increase the model capacity efficiently, the deep Transformer proposed in \cite{irie2020much} replaced the single-layer feed-forward network in each Transformer sub-layer by a deep neural network with residual connections. 

Training deep Transformers can be difficult as it often gets caught in a bad local optimum. Therefore, to enable training deep Transformer, iterated loss \cite{tjandra2019deja} is used in \cite{wang2020transformer}. It allows output of some intermediate transformer layers to calculate auxiliary cross entropy losses which are interpolated to configure the final loss function. Apart from that, ``gelu" (Gaussian error linear units) \cite{hendrycks2016gaussian} activation function is used in the feed-forward network of each Transformer layer. Out of the different explored approaches, positional embedding with a convolutional block before each Transformer layer has shown the best performance. 

A self-attention based ASR model has been proposed in \cite{sperber2018self} by replacing the pyramidal recurrent block of LAS model at the encoder side with multi-head self-attention block. As self-attention computes similarity of each pair of input frames, the memory grows quadratically with respect to the sequence length. To overcome this, authors have applied a downsampling to the sequence length before feeding it to every self-attention block.  This downsampling is done by reshaping the sequences and it is a trade-off between the sequence length and the dimension. If the sequence length is reduced by a factor $ a $, then the dimension increased by the same factor. Specifically, 
$ X \in \mathcal{R}^{l \times d}  \underbrace{\rightarrow}_{reshape}  \hat{X} \in \mathcal{R}^{\frac{l}{a} \times {ad}}  $. Therefore, memory consumption to compute the attention matrices is reduced by $ a^2 $. Unlike in \cite{vaswani2017attention} where position information is added to input sequence before feeding to the self-attention block, in \cite{sperber2018self}, authors have claimed that adding positional information to the acoustic sequence makes the model difficult to read content. Therefore, position information is concatenated to the acoustic sequence representation and this concatenated sequence is passed to the self-attention blocks. In addition, to enhance the context relevance while calculating the similarity between speech frames, a Gaussian diagonal mask with learnable variance is added to the attention heads. Specifically, an additional bias matrix is added to Equation \eqref{eq:self-attention} as given by \eqref{eq:self-attention-gaussian}.

\begin{equation}
Attention(Q,K,V) = Softmax(\frac{QK^T}{\sqrt(d_k)}+M)V
\label{eq:self-attention-gaussian}, 
\end{equation} where $ M $ is matrix whose values around the diagonal are set to a higher value to force the self-attention attending  in a local range around each speech frame. The elements of this matrix are calculated by a Gaussian function: $ M_{i,j} = \frac{-(j-k)^2}{2\sigma^2} $, $ \sigma $ is a learnable parameter.  

The quadratic computation complexity during the self-attention computation using \eqref{eq:self-attention} has been reduced down to linear in \cite{katharopoulos2020transformers} where the authors have proposed to use the dot product of kernel feature maps for the similarity calculation between the speech frames  followed by the use of associative property of matrix products. 

For better incorporating long-term dependency using Transformers, in \cite{dai2019transformer} Transformer-XL was proposed for machine-translation. In Transformer-XL, a segment-level recurrence mechanism is introduced which enables the reuse of past encoder states (output of the previous layers) at the training time to maintain a longer history of contexts until they become sufficiently old. Therefore, queries at current layer have access to the key-value pairs of current layer as well as previous layers.  Based on this concept, Compressive Transformer \cite{rae2019compressive} was proposed and it was applied to ASR to effectively incorporate long-term dependencies. In \cite{rae2019compressive}, instead of discarding older encoder states, they were preserved in a compressed form. \cite{irie2020much} also explored sharing previous encoder states but reused only key vectors from previous layers.  

Another Transformer-based ASR model is proposed in \cite{zhang2020transformer} as an adaptation of RNN-Transducer based model \cite{rao2017exploring} which uses two RNN-based encoders for audio and labels respectively to learn the alignment between them. In \cite{zhang2020transformer}, audio and label encoders are designed with Transformer networks. Given the previous predicted label from the target label space, the two encoder outputs are combined by a joint network.

Vanilla Transformer and the deep Transformer models have a number of layers stacked in both encoder and decoder sides. Each layers and their sub-layers have their own parameters and processing them is computationally expensive. In \cite{li2019improving}, a parameter sharing approach has been proposed for Transformer network. The parameters are initialised at the first encoder and decoder layers and thereafter, re-used in the other layers. If the number of encoder and decoder layers is $ N $ and the total number of parameters in each layer 
is $ M $, then instead of using $ N \times M $ parameters in both encoder and decoder sides, in \cite{li2019improving} only $ M $ parameters are used. There is a performance degradation due to sharing the parameters. To overcome that, speech attributes such as, duration of the utterance, sex and age of the speaker are augmented with the ground truth labels during training.


In self-attention based Transformer models, each speech frame attends all other speech frames of the entire sequence or within a window. However, some of them like frames representing $ silence $ are not crucial for modelling long-range dependencies and may present multiple times in the attended sequence. Therefore, these frames should be avoided. The attention weights (or probabilities) are obtained using $ softmax $ function which generates non-zero probabilities and therefore, insignificant frames are also assigned to some attention weights. To overcome this, in \cite{shi2020weak} weak-attention suppression (WAS) mechanism is proposed. WAS induced sparsity over the attention probability distribution by setting attention probabilities to zero which are smaller than a dynamically determined threshold. More specifically, the threshold is determined by \eqref{eq:WAS}. After that, the rest non-zero probabilities are re-normalised by passing through a $ softmax $ function.

\begin{equation}
\theta_i = m_i - \gamma_i \sigma_i,
\label{eq:WAS}  
\end{equation} where $ \theta_i $ is the threshold, $ m_i $ and $ \sigma_i $ are the mean and standard deviation of the attention probability for the $ i^{th} $ frame in the query sequence. $ \gamma $ is a scaling factor which ranges from $ 0 $ to $ 1 $ and experimentally, $ 0.5 $ provided the best result.

\section{Streaming Speech Recognition}

\label{sec:streaming}

 For offline speech recognition, the entire speech frames are already available before the transcription starts. However, for streaming environment, it is not possible to pass the entire speech through the encoder before the prediction starts. Therefore, to transcribe streaming speech, attention mechanism mostly focuses on a range or a window of input speech frames. Specifically, streaming spech recognition relies on local attention. In this section, we will discuss the development of attention models for streaming speech recognition. This section is divided into two sub-sections to explore RNN- and Transformer-based literature. 


\subsection{RNN-based models}

In this section, we will discuss the literature where attention mechanism is applied for streaming speech recognition with RNN-based encoder decoder models. To work with streaming speech, it is first required to obtain the speech frame or the set of speech frames on which attention mechanism will work. A Gaussian prediction-based attention mechanism is proposed in \cite{hou2017gaussian} for streaming speech recognition. Instead of looking at the entire encoder hidden states, at each decoder time step, only a set of encoder hidden states are attended based on a Gaussian window. The centre and the size of window at a particular decoder time step, $ t $  are determined by its mean ($ \mu_t $) and variance ($ \sigma_t $) which are predicted given the previous decoder state. Specifically, the current window centre is determined by a predicted moving forward increment ($ \triangle \mu_t $) and last window centre.  $ \mu_t = \triangle \mu_t + \mu_{t-1} $. A different approach compared to \eqref{eq:matchg} has been considered to calculate the similarity between $ j^{th} $ encoder state (within the current window) and $ i^{th} $ encoder state and it is given by \eqref{eq:gaussain-attention}:

\begin{equation}
e_{i,j} = exp(-\frac{(i-\mu_t)^2}{2\sigma_t^2})
\label{eq:gaussain-attention}.
\end{equation}

A hard monotonic attention mechanism is proposed in \cite{raffel2017online}. Only a single encoder hidden state $ h_i $ ($ i $ represents a decoder time step and $ h_i $ represents the only encoder state selected for output prediction at $ i^{th} $ decoder time step) which scores the highest similarity with the current decoder state is selected by passing the concerned attention probabilities through a categorical function. A stochastic process is used to enable attending encoder hidden states only from left to right direction. At each decoder time step, the attention mechanism starts processing from $ h_{i-1} $ to the proceeding states. $ h_{i-1} $ is the encoder state which was attended at last decoder time step. Each calculated similarity score ($ e_{i,j} $) is then sequentially passed through a logistic sigmoid function to produce selection probabilities ($ p_{i,j} $) followed by a Bernoulli distribution and once it outputs $ 1 $, the attention process stops. The last attended encoder hidden state, $ h_i $ at the current decoder time step is then set as the context for the current decoder time step, i.e. $ c_i = h_i $. Although the encoder states within the window of boundary $ [h_{i-1},h_i] $ are processed, only a single encoder state is finally selected for the current prediction.

 \cite{raffel2017online} provides linear time complexity and online speech decoding, it only attends a single encoer state for each output prediction and it may cause degradation to the performance. Therefore, monotonic chunkwise attention (MoChA) is proposed in \cite{chiu2018monotonic} where decoder attends small ``chunks" of encoder states within a window containing a fixed number of encoder states prior to and including $ h_i $. Due to its effectiveness, MoChA is also used to develop an on-device commercialised ASR system \cite{kim2019attention}. To increase the effectiveness of the matching scores obtained to calculate the attention probabilities between the decoder state and the chunk encoder states, multi-head monotonic chunkwise attention (MTH-MoChA) is proposed in \cite{liu2020multi}. MTH-MoChA splits the encoder and decoder hidden states into $ K $ heads. $ K $ is experimentally set as $ 4 $. For each head, matching scores, attention probabilities and the context vectors are calculated to extract the dependencies between the encoder and decoder hidden states. Finally, the average context vector over all the heads takes part in decoding.

The pronunciation rate among different speakers may vary and therefore, the attention calculated over the fixed chunk size may not be effective. To overcome this, in \cite{fan2019online} an adaptive monotonic chunkwise attention (AMoChA) was proposed where attention at current decoder time step is computed over a window whose boundary $ [h_{i-1},h_i] $ is computed as in \cite{raffel2017online}. Within the window, whichever encoder states results in $ p_{i,j} > 0.5 $ or $ e_{i,j} > 0 $ are attended. Hence, the chunk size is adaptive instead of constant.

The input sequence or the encoder states of length $ L $ is divided equally into $ W $ in  \cite{jaitly2016online}. So, each block contains $ B = \frac{L}{W} $ encoder states, while the last block may contain fewer than B encoder states. In this model, each block is responsible for a set of output predictions and attention is computed over only the concerned blocks and not the entire encoder states. Once the model has finished attending all the encoder states of a block and predicting the required outputs, it emits a special symbol called $ <epsilon> $ which marks the end of the corresponding block processing and the model proceeds to attend the next block. The effectiveness of this model has been enhanced in \cite{sainath2018improving} by extending the attention span. Specifically, the attention mechanism looks at not only the current block but the $ k $ previous blocks. Experimentally, $ k $ is set as $ 20 $.

The authors of \cite{inaguma2020minimum} have identified the latency issue in streaming attention-based models. In most streaming models, the encoder states are attended based on a local window. Computing the precise boundaries of these local windows is a computational expensive process which in turn causes a delay in the speech-to-text conversion. To overcome this issue, in \cite{inaguma2020minimum} external hard alignments obtained from a hybrid ASR system is used for frame-wise supervision to force the MoChA model to learn accurate boundaries and alignments. In \cite{liu2020low} performance latency is reduced by proposing a unidirectional encoder with no future dependency. Since each position does not depend on future context, the decoder hidden states are not required to be re-computed every time a new input chunk arrives and therefore, the overall delay is reduced. 

In \cite{wang2020attention}, attention mechanism has been incorporated in RNN-Transducer (RNN-T) \cite{graves2012sequence,graves2013speech} to make streaming speech recognition more effective and efficient. RNN-T consists of three sections: (i) a RNN encoder which processes an input sequence to encoder hidden states, (ii) a RNN decoder which is analogues to a language model takes the previous predicted symbol as input and outputs decoder hidden states, and (iii) a joint network that takes encoder and decoder hidden states at the current time step to compute output logit which is responsible to predict the output symbol when passed through a softmax layer. In \cite{wang2020attention}, at the encoder side, to learn contextual dependency, a multi-head self-attention layer is added on the top of RNN layers. In addition, the joint network attends a chunk of encoder hidden states instead of attending only the current hidden state at each time step.

LAS model is primarily proposed for offline speech recognition. However, it has been modified with silence modelling for working in the streaming environment in \cite{hsiao2020online}. Given streamable encoder and a suitable attention mechanism (hard monotonic, chunkwise or local window-based instead of global), the main limitation of LAS model to perform in streaming environment is a long enough silence between the utterances to make decoder believe it is the end of speech. Therefore, the LAS decoder terminates the transcription process while the speaker is still active (i.e. early stopping). This limitation is addressed in \cite{hsiao2020online} by incorporating reference silence tokens during the training phase to supervise the model when to output a silence token instead of terminating the process during the 
inference phase.

\subsection{RNN-free Transformer-based models}

In this section, we will discuss the literature where RNN-free self-attention models are used for streaming speech recognition. Self-attention aligner \cite{dong2019self} which is designed based on the Transformer model proposes a chunk hoping mechanism to provide support to online speech recognition. Transformer-based network requires the entire sequence to be obtained before the prediction starts and hence, not suitable for online speech recognition. In \cite{dong2019self}, the entire sequence is partitioned into several overlapped chunks, each of which contains three parts belonging to current, past and future. Speech frames or encoder states of the current part are attended to provide the output predictions belonging to the corresponding chunk. The past and future parts provide contexts to the identification of the current part. After attending a chunk, the mechanism hops to a new chunk to attend. The number of speech frames or encoder states hopped between two chunks is same as the current part of each chunk. A similar method was proposed in augmented memory Transformer \cite{wu2020streaming} where an augmented memory bank is included apart from partitioning the input speech sequence. The augmented memory bank is used for carrying the information over the chunks, specifically by extracting key-value pairs from the projection of concatenated augmented memory bank and the relevant chunk (including past, current and future parts).

Transformer transducer model \cite{yeh2019transformer} uses truncated self-attention to support streaming ASR. Instead of attending the entire speech sequence at each time step $ t $, truncated self-attention mechanism allows attending speech frames within the window of [$ t-L, t+R $] frames. $ L $ and $ R $ represent the frame limits to the left and right respectively. In \cite{yeh2019transformer}, positional encoding in input embedding is done by causal convolution \cite{mohamed2019transformers} to support online ASR. In another variation of Transformer transducer \cite{zhang2020transformer}, the model restricts attending to the left side of the current frame only by masking the attention scores to the right of the current frame. The attention span is further restricted by attending the frames within a fixed-size window at each time step. 

A chunk-flow mechanism is proposed in \cite{tian2019self} to support streaming speech recognition in self-attention based transducer model. The chunk-flow mechanism restricts the span of self-attention to a fixed length chunk instead the whole input sequence. The fixed length chunk proceeds along time over the input sequence. Not attending the entire input sequence may degrade the performance. However, it is still kept satisfactory by using multiple self-attention heads to model longer dependencies. The chunk-flow mechanism at time $ t $ for the attention head $ h_i $ is given by \eqref{eq:chunk-flow}

\begin{equation}
h_{i,t}= \sum_{\tau = t-N_l}^{t+N_r} \alpha_{i,\tau}s_{\tau}
\label{eq:chunk-flow},
\end{equation} where $ N_l $ and $ N_r $ represent the number of speech frames to the left and right of the current time $ t $. $ N_l $ and $ N_r $ determine the chunk span and experimentally they are chosen as 20 and 10 respectively. $ s_{\tau} $ represents the $ \tau^{th} $ vector in the input sequence and $ \alpha_{i,\tau} = Attention(s_{\tau}, K,V) $; $ K = V = chunk_{\tau} $


A streaming friendly self-attention mechanism, named as time-restricted self-attention is proposed in \cite{povey2018time}.
It works by restricting the speech frame at current time step to attend only a fixed number of frames to its left and right and thus it does not allow attending  each speech frame to attend all other speech frames. Experimentally, these numbers are set to 15 and 6 for left and right sides, respectively. Similarly, in \cite{wang2020transformer}, each Transformer layer is restricted to attend a fixed limited right context during inference. A special position embedding approach also has been proposed by adding a one-hot encoder vector with the value vectors. The one-hot encoder vector consists of all zeros except a single one corresponding to the attending time step with respect to all the time steps in the current attention span. This mechanism is also used in the encoder side of streaming transformer model \cite{moritz2020streaming}.


Synchronous Transformer \cite{tian2020synchronous} is proposed to support streamable speech recognition using self-attention mechanism to overcome the requirement of processing all speech frames before decoding starts. While calculating the self-attention, every speech frame is restricted to process only the frames left to it and ignore the right side. Also, at the decoder time step, encoded speech frames are processed chunkwise. The encoded speech frames are divided into overlapped chunks to maintain the smooth transition of information between chunks. At each decoder time step, the decoder predicts an output based on the last predicted output and the attention calculated over the frames belonging to a chunk only and therefore, avoids attending the entire speech sequence.  

To make Transformer streamable, chunk self-attention encoder and monotonic truncated attention-based self-attention decoder is proposed in \cite{miao2020transformer}. At the encoder side, the input speech is split into isolated chunks of fixed length inspired by MoChA. At the decoder side, encoder-decoder attention mechanism \cite{vaswani2017attention}  is replaced by truncated attention \cite{miao2019online}.  The encoder embedding is truncated in a monotonic left to right approach and then attention applied over the trunacted outputs. After that, the model is optimised by online joint CTC-attention method \cite{miao2019online}.

Monotonic multihead attention (MMA) is proposed in \cite{ma2019monotonic} to enable online decoding  in Transformer network by replacing each encoder-decoder attention head with  a monotonic attention (MA) head. Each MA head needs to be activated to predict a output symbol. If any MA head failed or delayed to learn alignments, it causes delay during inference. The authors of \cite{inaguma2020enhancing} have found that only few MA heads (dominant ones) learn alignments effectively and others do not. To prevent this and to let each head learning alignments effectively, HeadDrop regularisation is proposed. It entirely masks a part of the heads at random and forces the rest of non-masked heads to learn alignment effectively. In addition, the redundant MA heads are pruned in the lower layers to further improve the team work among the attention heads. Since MA is hard attention, chunkwise attention is applied on the top of each MA head to enhance the quality of context information.

\section{Conclusion}

\label{sec:conclusion}

In this survey, how different types of attention models have been successfully applied to build automatic speech recognition models is presented. We have discussed various approaches to deploy attention model into the RNN-based encoder-decoder framework. We have also discussed how self-attention replaces the need of recurrence and can build effective and efficient ASR models. Speech recognition can be performed offline as well as online and in this paper, we have discussed various aspects of the offline and online ASR development.

\balance

\bibliographystyle{IEEEtran}
\bibliography{speechref}	
	
\end{document}